\documentclass[10pt,showpacs,notitlepage,tightenlines,eqsecnum,floats,aps,amsmath,amssymb,nofootinbib,prd,twocolumn]{revtex4-1}
\usepackage[latin9]{inputenc}
\usepackage{color}
\usepackage{amsmath}
\usepackage{amssymb}
\usepackage{graphicx}
\usepackage{esint}
 \usepackage{breqn}
\makeatletter
\usepackage[unicode=true,
 bookmarks=false,
 breaklinks=false,pdfborder={0 0 1},backref=section,colorlinks=true]
 {hyperref}
\hypersetup{linkcolor=blue,urlcolor=black,citecolor=blue}

\providecommand{\tabularnewline}{\\}

\usepackage[usenames,dvipsnames]{xcolor}

\makeatother

\begin{document}

\title{Highly nonlinear wave solutions in a dual to the chiral model}

\author{S. G. Rajeev}

\email{s.g.rajeev@rochester.edu}

\altaffiliation[Also at the ]{Department of Mathematics}

\author{Evan Ranken}

\email{evan@pas.rochester.edu}

\affiliation{Department of Physics and Astronomy\\ 
University of Rochester \\ 
Rochester$,$ New York 14627$,$ USA}

\date{\today}
\begin{abstract}
We consider a two-dimensional scalar field theory with a nilpotent
current algebra, which is dual to the Principal Chiral Model. The
quantum theory is renormalizable and not asymptotically free: the
theory is strongly coupled at short distances (encountering a Landau
pole). We suggest it can serve as a toy model for $\lambda\phi^{4}$
theory in four dimensions, just as the principal chiral model is a
useful toy model for Yang-Mills theory. We find some classical wave
solutions that survive the strong coupling limit and quantize them
by the collective variable method. They describe excitations with
an unusual dispersion relation $\omega\propto|k|^{\frac{2}{3}}$ .
Perhaps they are the ``preons'' at strong coupling, whose bound states
form massless particles over long distances.
\end{abstract}
\pacs{11.10.Kk, 42.65.-k, 03.50Kk, 03.65.Fd }
\maketitle

\section{Introduction}

We study  the field theory \cite{Nappi1980,NappiWitten1993,CurtrightZachos} with equations
of motion
\begin{equation}
\ddot{\phi}=\lambda\left[\dot{\phi},\phi'\right]+\phi'',\label{eq:FieldEqn}
\end{equation}
where $\phi$ is valued in a Lie algebra, $\phi:\mathbb{R}^{1,1}\to \mathfrak{su}(2)$.
This follows from the action
\begin{dmath}
S_{1}\equiv\int\mathcal{L}_{1}dxdt=\int\mathrm{Tr}\left\{ \frac{1}{2\lambda}\dot{\phi}^{2}-\frac{1}{2\lambda}{\phi'}^{2}+\frac{1}{3}\phi[\dot{\phi},\phi']\right\} dxdt.\label{eq:s1}
\end{dmath}
In the $\lambda\rightarrow0$ limit, these equations admit linear
wave solutions. But in the high-coupling regime, the theory is dominated
by nonlinear effects. 

$S_{1}$ is closely tied to other models and subjects, which we elaborate
on in section \ref{sec:relations}. These include the study of slow
light, the Wess-Zumino-Witten (WZW) model, and the mathematical theory
of hypoelliptic operators. Of particular interest in this paper, the
model described by $S_{1}$ is also classically dual to the well-studied
principal chiral model, described by the action
\begin{dmath}
S_{2}=\int\mathcal{L}_{2}dxdt=\frac{1}{2f}\int\mathrm{Tr}\ \left\{ \left(g^{-1}\dot{g}\right)^{2}-c^{2}\left(g^{-1}g'\right)^{2}\right\} dxdt
\end{dmath}
where $g:\mathbb{R}^{1,1}\to SU(2)$. This is a special case of the
nonlinear sigma model, with target space $SU(2)$. 

Despite their classical equivalence, $S_{1}$ and $S_{2}$ lead to
entirely different quantum theories. $S_{2}$ gives an asymptotically
free theory: at short distances $f\to0$, giving us free massless
excitations. But the true particles that survive to long distances
are bound states of non-zero mass \cite{PCM,PolyakovWiegmann}. For this reason, the
principal chiral model is often used as a toy model for 4-dimensional
Yang-Mills theory, notorious for its mathematical complexity. Not
only do the two theories share similar short-distance behavior, but
the existence of a mass gap in the principal chiral model has served as a proof of
concept for the conjectured mass gap in Yang-Mills (though neither can yet be proven with full mathematical rigor).

$S_{1}$, on the other hand, leads to a renormalizable but not asymptotically
free quantum theory. At short distances the coupling constant $\lambda\to\infty$,
while at long distances we have weakly nonlinear massless excitations.
It makes sense to use $S_{1}$ as a 2-D toy model for strongly coupled
theories, in particular 4-dimensional $\lambda\phi^{4}$ theory\footnote{that is, pure $\lambda\phi^{4}$ theory, describing a Higgs-like particle
with no coupling to fermions}. The behavior of quantum field theories at high coupling is notoriously
intractable, and the physical meaning of such theories is still
up for debate. For this reason, it is still necessary to search for
simple examples of such theories and try to gleam what meaning, if
any, they have in the short distance limit.

In addition to sharing short-distance behavior, both the $S_{1}$
model and $\lambda\phi^{4}$ theory can be described by hypoelliptic hamiltonian operators
with a step-3 nilpotent bracket algebra, suggesting some algebraic
structure in common (section \ref{sub} and appendix \ref{sec:Appendix:nilpotent}). The $S_{1}$ model's relative simplicity makes
it a good candidate for attempting to probe the high coupling regime
of field theories in general, but the connection to $\lambda\phi^{4}$
theory seems the closest. Additionaly, its classical duality to the
principal chiral model motivates a juxtaposition of the two theories
in the classical and quantum formulations. 

To glimpse what becomes of our theory in the high coupling limit,
we take the modest approach of finding nonlinear wave-type solutions
to the classical model which survive the $\lambda\rightarrow\infty$
limit (section \ref{sec:Reduction-mech}). This set of solutions defines
a mechanical system or ``reduced system'' in each of the dual models.
While they physically appear very different, the resulting classical
solutions can be mapped from one system to another. We quantize these
collective variables to determine their dispersion relation (section
\ref{sec:reduced_system_quant}) in the short distance limit for each
theory. We have in mind the sine-Gordon theory, whose solitons turn
out to be the fundamental constituents which bind to form the scalar
particles \cite{sineGordon,QuantThSolitons}.

These reduced quantum theories yield two different results. In particular,
the reduced model of $S_{1}$ has an exotic dispersion relation in
the short distance limit. We postulate that its spectrum may hint
at the fundamental constituents of the highly coupled theory, which
need not behave like traditional particles at all. In section \ref{sec:Conclusion-and-future}
we offer concluding remarks and a side-by-side comparison of our work
with $S_{1}$ and $S_{2}$.

\subsection*{On The Notation}

We regard $\phi=\frac{1}{2i}\left[\phi_{1}\sigma_{1}+\phi_{2}\sigma_{2}+\phi_{3}\sigma_{3}\right]=\frac{1}{2i}\boldsymbol{\phi}\cdot\boldsymbol{\sigma}$
as a traceless anti-hermitian matrix. Recall then that the commutator
and cross product are related by
\begin{equation}
[X,Y]=\frac{1}{2i}\left(\mathbf{X}\times\mathbf{Y}\right)\cdot\boldsymbol{\sigma}.
\end{equation}
Also, we define $\mathrm{Tr}X\equiv-2\mathrm{tr}X$ so that 
\begin{equation}
\mathrm{Tr}\phi^{2}=\phi_{1}^{2}+\phi_{2}^{2}+\phi_{3}^{2}.
\end{equation}
In relativistically invariant notation, (\ref{eq:FieldEqn}) and (\ref{eq:s1})
can be written as
\begin{equation}
\partial^{\mu}\partial_{\mu}\phi_{a}-\frac{\lambda}{2}\ \epsilon_{abc}\epsilon^{\mu\nu}\partial_{\mu}\phi^{b}\partial_{\nu}\phi^{c}=0,
\end{equation}
\begin{dmath}
S_{1}=\frac{1}{2\lambda}\int\partial_{\mu}\phi^{a}\partial_{\nu}\phi^{a}\eta^{\mu\nu}d^{2}x+\frac{1}{6}\int\epsilon_{abc}\phi^{a}\partial_{\mu}\phi^{b}\partial_{\nu}\phi^{c}\epsilon^{\mu\nu}d^{2}x
\end{dmath}
where $\mu,\nu=0,1$ and $a,b,c=1,2,3$ ; also, $\epsilon^{\mu\nu},\epsilon_{abc}$
are the Levi-Civita tensors. This is a particular case of the general
sigma model studied in \cite{SigmaModelString} as the background
of string theory, with a flat metric on the target space and a constant
3-form field $\epsilon_{abc}$.

\section{relation to other models \label{sec:relations}}

\subsection{The $c\rightarrow0$ Limit and Slow Light\label{sub:Relation-to-Slow}}

Consider the equations of motion (\ref{eq:FieldEqn}) where the speed
of linear propagation at low coupling is taken to be $c$ rather than
1: 
\begin{equation}
\ddot{\phi}=\lambda\left[\dot{\phi},\phi'\right]+c^{2}\phi''.\label{eq:field_eqn_c}
\end{equation}
If we rescale $\phi\to\lambda^{a}\phi$, $t\to\lambda^{b}t$, this
becomes 
\begin{equation}
\lambda^{a-2b}\ddot{\phi}=\lambda^{1+2a-b}\dot{\phi}\times\phi'+c^{2}\lambda^{a}\phi''.
\end{equation}
Set $a=2b$ and $1+2a=b$ to get
\begin{equation}
\ddot{\phi}=\dot{\phi}\times\phi'+c^{2}\lambda^{-\frac{2}{3}}\phi''.
\end{equation}
Thus the strong coupling limit $\lambda\to\infty$ at fixed $c$ is
equivalent to the limit $c\to0$ with $\lambda=1$:
\begin{equation}
\ddot{\phi}=\dot{\phi}\times\phi'.\label{eq:StrongCouplingEqn}
\end{equation}

The strongly coupled limit can be thought of as the limit in which
the waves move very slowly. It has been noted in that literature \cite{SlowLight}
that when the speed of light in a medium is small, nonlinear effects
are magnified. Although the specific equations appearing there are
different, it is possible that the solutions of the sort we study
are of interest in that context as well.

From a field theoretic context, the equivalence of these limits seems
troubling. At short distances, the highly-coupled theory will not
be relativistic. It is a sort of ``post-relativistic'' regime, where
$c\rightarrow0$. This is much the opposite of the case in the theory
of $S_{2}$; there the short-distance excitations are massless, but form massive bounds states which survive to long distances and can be non-relativistic in the
traditional $c\rightarrow\infty$ sense. Perhaps some exotic
excitations at high coupling are in fact the fundamental constituents in the $S_1$-model,
forming as bound states the ordinary massless particles which appear
in the long distance limit. As we know from the quark model, the short
distance excitations do not need to be particles in the usual sense;
they could be confined. In any case, it is important to know what solutions
might survive the high coupling limit, whether they be unphysical
or simply unintuitive.

We will see an example of wave solutions which classically survive
the $c\rightarrow0$ limit, continuing to propagate through nonlinearity
alone. Since the energy density is constant, these solutions do not
violate causality: they are analogous to the Continuous Wave solutions
in a medium where the phase velocity is greater than $c$.

\subsection{Sub-Riemannian Geometry and the Strong Coupling Limit \label{sub}}

Many physical problems (Yang-Mills, Fluid Mechanics) become intractable in the strong coupling limit where the non-linearities dominate. It would be nice to have a unified geometric approach to understanding these systems. We have such an approach  in the weak coupling limit:  small perturbations around a stable equilibrium are equivalent to a harmonic oscillator. 

A larger picture emerges if we think in terms of Riemannian and sub-Riemannian geometry.  
The orbits of many mechanical systems of physical interest (again, Yang-Mills or incompressible Fluids) can be thought of as geodesics in some appropriate Riemannian manifold. In the simplest case, the harmonic oscillator describes geodesics in the Heisenberg group. The anharmonic oscillator (and many nonlinear field theories with quartic coupling) can also be thought of as geodesic motion on a nilpotent Lie group, by introducing an additional generator (see Appendix \ref{sec:Appendix:nilpotent} for  more detail).

 In the limit of strong coupling, the  metric degenerates and becomes sub-Riemannian \cite{subRiemannian}.  
That is, the contravariant metric tensor has some zero eigenvalues so that it can
be written as $\sum_{j}X_{j}\otimes X_{j}$ for some vector fields
$X_{j}$ whose linear span may be smaller than the tangent space.
Moreover, in the cases of interest, these vector fields satisfy the celebrated  H{\"o}rmander condition: $X_{j}$ along with their repeated commutators span the tangent spaces at every point.
In such a case, there are still  geodesics connecting every pair of sufficiently close
points (Chow-Rashevskii theorem, \cite{subRiemannian}). Thus, we can define a distance between pairs of points as the shortest length of geodesics.

These ideas came to the notice of many physicists following a model for the self-propulsion
of an amoeba \cite{ShapereWilczek}, though they have roots in the Carnot-Caratheodory
geometric formalism of thermodynamics and in control theory. H{\"o}rmander
\cite{Hormander} discovered independently that the same criterion
is sufficient for the sub-Riemannian Laplace operator $\Delta=\sum_{j}X_{j}^{2}$
to be \emph{hypoelliptic}, meaning the solution $f$ to the inhomogenous
equation $\Delta f=u$ is smooth whenever the source $u$ is smooth. This
can be thought of the quantum version of the above condition on subgeodesic
connectivity. 

This kind of sub-Riemannian geometry may present a powerful geometric framework for strongly-coupled field theories. The example we work out in this paper is arguably the simplest interesting case of a strongly coupled field theory, and the solutions we study correspond to sub-Riemannian geodesics in the limit $\lambda\to \infty$. We hope to apply such geometric ideas to other cases in the future, using this as a prototype.

\subsection{Relation to the WZW model }

We can also regard our equations as a limiting case of the Wess-Zumino-Witten
model\footnote{Witten's Tr is our tr. His $\lambda$ is our $\lambda_{1}$.}
\cite{Witten1984}
\begin{dmath}
S_{WZW}=\frac{1}{4\lambda_{1}^{2}}\int\mathrm{tr}\ \partial^{\mu}g\partial_{\mu}g^{-1}d^{2}x+\frac{n}{24\pi}\int_{M_{3}}\mathrm{tr}\ (g^{-1}dg)^{3}
\end{dmath}
as $n\to\infty$ and $\lambda_{1}\to0$, keeping $\lambda=\lambda_{1}^{2}\left(n/2\pi\right)^{\frac{2}{3}}$
fixed\footnote{Here, $M_{3}$ is a 3-manifold of which the two-dimensional space-time
is the boundary. We do not require $\lambda_{1}^{2}$ to take the
conformally invariant value $\frac{4\pi}{n}$ .}. To see this, let $g(x)=e^{bi\sigma_{a}\phi^{a}(x)}$ and expand
in powers of $b$:
\begin{dmath}
S_{WZW}=\frac{b^{2}}{2\lambda_{1}^{2}}\int\partial^{\mu}\phi^{a}\partial_{\mu}\phi^{a}+\frac{n}{24\pi}b^{3}\int_{M_{3}}2\epsilon_{abc}\ d\phi^{a}d\phi^{b}d\phi^{c}\ +\cdots
\end{dmath}
To this order the WZW term is an exact differential, so we can write
it as an integral over space-time
\begin{dmath}
S_{WZW}=\frac{1}{2}\frac{b^{2}}{\lambda_{1}^{2}}\int\partial^{\mu}\phi^{a}\partial_{\mu}\phi^{a}+\frac{n}{12\pi}b^{3}\int\epsilon_{abc}\ \phi^{a}d\phi^{b}d\phi^{c}\ +\cdots
\end{dmath}

$S_{WZW}$ reduces to $S_{1}$ if we identify $b^{3}=2\pi/n$ and
$\lambda$ as above. By taking this limit, we can easily get the renormalization
of our model. Recall that \cite{Witten1984} the one loop renormalization
group equation of the $O(N)$ WZW model is
\begin{equation}
\frac{d\lambda_{1}^{2}}{d\log\Lambda}=-\frac{\lambda_{1}^{4}(N-2)}{2\pi}\left[1-\left(\frac{\lambda_{1}^{2}n}{4\pi}\right)^{2}\right]
\end{equation}
We need the particular case of $N=4$ corresponding to the target
space being $S^{3}\approx SU(2)$. Thus in our limit $n\to\infty$,
$\lambda_{1}\to0$ keeping $\lambda$ fixed,
\begin{equation}
\frac{d\lambda}{d\log\Lambda}=\frac{\lambda^{4}}{4\pi}
\end{equation}
It is useful to take this limit rather than calculating loop corrections
from scratch, as the renormalization group evolution of the WZW has
been studied to high order \cite{Bos,Xi}. Including these higher
order terms does not alter the short-distance divergence of $\lambda$.

\subsection{Duality with the principal chiral model}

We have now seen that the $S_{1}$ model is strongly coupled in the short-distance
limit. Yet, as a classical field theory, it can be viewed \cite{Nappi1980,NappiWitten1993}
as a dual to the asymptotically free principal chiral model with equation
of motion
\begin{equation}
\partial^{\mu}[g^{-1}\partial_{\mu}g]=0,\quad\;\; g:\mathbb{R}^{1,1}\to SU(2).
\end{equation}
To see this, we define the currents
\begin{equation}
I=\frac{1}{\lambda}\dot{\phi},\quad J=\phi'
\end{equation}
so the equations of motion become
\begin{equation}
\dot{J}=\lambda I',\quad\dot{I}=\lambda[I,J]+\frac{1}{\lambda}J'.\label{eq:eqmo}
\end{equation}
We can solve the second equation with the relations
\begin{equation}
I=\frac{1}{\lambda^{2}}g^{-1}g',\quad J=\frac{1}{\lambda}g^{-1}\dot{g}.\label{eq:dual}
\end{equation}

Then the first equation becomes
\begin{equation}
\partial_{0}\left[g^{-1}\dot{g}\right]=\partial_{1}\left[g^{-1}g'\right]
\end{equation}
which is the non-linear sigma model. Thus, the same classical equations
of motion follow from the action
\begin{equation}
S_{2}=\frac{1}{2f}\int\mathrm{Tr}\ \left\{ \left(g^{-1}\dot{g}\right)^{2}-c^{2}\left(g^{-1}g'\right)^{2}\right\} dxdt
\end{equation}
if we identify $f=\lambda^{2}$. A summary of relevant correspondences
in the dual models can be found in table \ref{tab:thetable} at the end of section \ref{sec:Conclusion-and-future}.

We also briefly note that our theory is closely related to the sigma model on the Heisenberg group (see \cite{Baaquie2005}).

\section{Reduction to A Mechanical System \label{sec:Reduction-mech}}

We will look at propagating waves of the form 
\begin{dmath}[compact]
\phi(t,x)=e^{Kx}R(t)e^{-Kx}+mKx,\quad K=\frac{i}{2}k\left(\begin{array}{cc}
1 & 0\\
0 & -1
\end{array}\right)
\end{dmath}
for constants $k,m$. These solutions are equivariant under translations:
the ``potential'' $\phi$ changes by an internal rotation and a
constant shift under translation, while the currents only change only
by the internal rotation. Thus, the energy density is constant. They
are to be contrasted with soliton solutions, which have energy density
concentrated at the location of the soliton. They are more analogous
to the plane wave solutions of the wave equation, or a Continous Wave
(CW) laser beam. Moreover, the currents
\begin{equation}
I=\frac{1}{\lambda}e^{Kx}\dot{R}e^{-Kx},\quad\; J=e^{Kx}\left\{ [K,R]+mK\right\} e^{-Kx}
\end{equation}
are periodic in space with wavelength $\frac{2\pi}{k}$. Defining
\begin{equation}
L\equiv[K,R]+mK,\quad S\equiv\dot{R}+\frac{1}{\lambda}K,
\end{equation}
We can write the equations of motion and identity (\ref{eq:eqmo})
in a symmetric form
\begin{equation}
\dot{L}=[K,S],\quad\dot{S}=\lambda[S,L].\label{eq:eqmo_ansatz}
\end{equation}
This new choice of variables will allow us to connect to the dual
theory, identify the conserved quantities and to pass to the quantum
theory more easily.

\subsection{The reduced system Lagrangian}

Three conserved quantities follow immediately:
\begin{eqnarray}  \label{conserved}
s^{2}k^{2}&\equiv& \mathrm{Tr}\ S^{2} \nonumber \\ C_{1}k^{2}&\equiv &\mathrm{Tr}SL \\ \nonumber C_{2}k^{2}&\equiv& \mathrm{Tr}\left[\frac{1}{2}L^{2}-\frac{1}{\lambda}KS\right]. 
\end{eqnarray}
The quantity $s$ will be of importance in the dual picture, while
the other constants have less obvious roles there. Moreover, we have
the identity
\begin{equation}
\mathrm{Tr}KL=mk^{2}.
\end{equation}

Of the six independent variables in $S$ and $L$, only two remain
after taking into account these constants of motion. The dynamics
are described by the effective lagrangian density (dropping a total
time derivative and an overall factor of volume of space divided by
$\lambda$)
\begin{multline*}
\mathcal{L}_{1} =\mathrm{Tr}\biggl\{ {\frac{1}{2}\dot{R}^{2}+\frac{\lambda}{3}R\left[\dot{R},\ [K,R]+mK\right] }\\  -\frac{1}{2}\left([K,R]+mK\right)^{2}\biggr\} 
\end{multline*}
\begin{dmath}
\label{eq:EffectiveLagrangian} 
=\mathrm{Tr}\left\{ \frac{1}{2}(S-\frac{1}{\lambda}K)^{2}+\frac{\lambda}{3}R\left[S-\frac{1}{\lambda}K,\ L\right]-\frac{1}{2}L^{2}\right\} 
\end{dmath}
and hamiltonian density
\begin{equation}
H_{1}=\mathrm{Tr}\left[\frac{1}{2}\left(S-\frac{1}{\lambda}K\right)^{2}+\frac{1}{2}L^{2}\right],\label{eq:Ham_Ansatz}
\end{equation}

\subsection{Reduction to One Degree of Freedom}

It is useful to work with the first two components of $R$ as a single
complex variable. Defining $Z=R_{1}+iR_{2}$, we can write explicitly
\begin{equation}
L=\frac{k}{2}\left(\begin{array}{cc}
im & \bar{Z}\\
-Z & -im
\end{array}\right).
\end{equation}
To describe the third component, we define
\begin{equation}
u\equiv\frac{1}{k}\dot{R}_{3}-\frac{1}{\lambda},\label{eq:u_def}
\end{equation}
allowing us to write a similarly compact expression for $S$,
\begin{equation}
S=\frac{1}{2i}\left(\begin{array}{cc}
uk & \dot{\bar{Z}}\\
\dot{Z} & -uk
\end{array}\right).
\end{equation}

The three conserved quantities (\ref{conserved}) can now be written
in terms of $Z$ and $u$ as
\begin{eqnarray}
s^{2}k^{2} & = & u^{2}k^{2}+|\dot{Z}|^{2}\nonumber \\
C_{1}k^{2} & = & \frac{ik}{2}\left[\bar{Z}\dot{Z}-\dot{\bar{Z}}Z\right]-mk^{2}u\nonumber \\
C_{2}k^{2} & = & \frac{k^{2}}{2}\left(m^{2}+\frac{2u}{\lambda}+|Z|^{2}\right).\label{eq:conserved_2}
\end{eqnarray}
Using the identity
\begin{equation}
\left(\frac{d}{dt}|Z|^{2}\right)^{2}=4|Z|^{2}|\dot{Z}|^{2}+(\dot{\bar{Z}}Z-\bar{Z}\dot{Z})^{2},
\end{equation}
we can combine these three equations to eliminate $Z$ and yield an
ODE for $u(t)$,
\begin{dmath}
\dot{u}^{2}=k^{2}\lambda^{2}\left\{ \left[2C_{2}-m^{2}-\frac{2}{\lambda}u\right](s^{2}-u^{2})-\left[mu+C_{1}\right]^{2}\right\} .
\end{dmath}

\subsection{Solution in terms of elliptic functions}

The ODE for $u(t)$ describes an elliptic curve. Setting $u=av+b$,
we can pick the constants
\begin{equation}
a=\frac{2}{k^{2}\lambda},\quad b=\frac{C_{2}\lambda}{3}
\end{equation}
to bring our ODE to Weierstrass normal form in terms of $v$:
\begin{equation}
\dot{v}^{2}=4v^{3}-g_{2}v-g_{3}.\label{eq:weierstrass_eqn}
\end{equation}
The somewhat unsightly expressions for $g_{2}$ and $g_{3}$ can be
obtained by symbolic computation:
\begin{multline}
g_{2}  =  \frac{1}{3}k^{4}\lambda^{2}\left(3C_{1}\lambda m+C_{2}^{2}\lambda^{2}+3s^{2}\right)\\ 
\shoveleft{g_{3}  =  \frac{1}{108}k^{6}\lambda^{4}\bigl( 27C_{1}^{2}+18C_{1}C_{2}\lambda m+4C_{2}^{3}\lambda^{2}} \\ 
 -36C_{2}s+27m^{2}s^{2}\bigr)
\end{multline}
The solution to the Weierstrass differential equation (\ref{eq:weierstrass_eqn})
is then
\begin{equation}
v(t)=\wp(t+\alpha)\;\Longrightarrow\;u(t)=\frac{2}{k^{2}\lambda}\wp(t+\alpha)+\frac{C_{2}\lambda}{3},
\end{equation}
where $\wp$ is the Weierstrass $P$-function and $\alpha$ is a complex
constant determined by the initial conditions. We can most immediately
solve for $R_{3}(t)$. Recalling (\ref{eq:u_def}), we have
\begin{equation}
\dot{R}_{3}=\frac{2}{k\lambda}\wp(t+\alpha)+k\left(\frac{C_{2}\lambda}{3}+\frac{1}{\lambda}\right).
\end{equation}

In order for to obtain a sensible solution, $\wp(t+\alpha)$ must
be real and bounded. This requires $\mathrm{Im}(\alpha)=|\omega_{2}|,$
where $\omega_{2}$ is the imaginary half-period of the Weierstrass
$P$-function (which depends on the elliptic invariants $g_{2}$,
\textbf{$g_{3}$}). The real part of $\alpha$ merely shifts our solution
in time, so we can take $\alpha=\omega_{2}$ for simplicity.
Using the relationship
\begin{equation}
\int\wp(u)du=-\zeta(u),
\end{equation}
where $\zeta$ is the Weierstrass $\zeta$-function, and taking $R_{3}(0)=0$
gives the solution
\begin{equation}
R_{3}(t)=\frac{2}{k\lambda}\left[\zeta(\omega_{2})-\zeta(t+\omega_{2})\right]+\left(\frac{C_{2}\lambda}{3}+\frac{1}{\lambda}\right)kt.
\end{equation}

The solution for the other two components is found by making the substitution
$Z=re^{i\theta}$ in (\ref{eq:conserved_2}). Writing $|Z^{2}|=r^{2}$
quickly yields
\begin{equation}
r^{2}(t)=\frac{4}{3}C_{2}-m^{2}-\frac{4}{k^{2}\lambda^{2}}\wp(t+\omega_{2}).
\end{equation}
Note that the choice of $\mathrm{Re}(\alpha)=0$ we made earlier implies
that $t=0$ is a turning point of the radial variable, as $\wp'(\omega_{2})$
is necessarily $0$. It is useful to write 
\begin{equation}
r^{2}(t)=\frac{4}{k^{2}\lambda^{2}}[\wp(\Omega)-\wp(t+\omega_{2})],
\end{equation}
where 
\begin{equation}
\wp(\Omega)=k^{2}\lambda^{2}\left(\frac{C_{2}}{3}-\frac{m^{2}}{4}\right).
\end{equation}
Then we can use the identity
\begin{equation}
\wp(z)-\wp(\Omega)=-\frac{\sigma(z+\Omega)\sigma(z-\Omega)}{\sigma^{2}(z)\sigma^{2}(\Omega)},
\end{equation}
where $\sigma$ is the Weierstrass $\sigma$-function, in order to
simplify a later result. We obtain the solution
\begin{equation}
r(t)=\frac{2}{\lambda k\sigma(\Omega)}\frac{\sqrt{\sigma(t+\omega_{2}+\Omega)\sigma(t+\omega_{2}-\Omega)}}{\sigma(t+\omega_{2})}
\end{equation}

To find $\theta(t)$ from (\ref{eq:conserved_2}), we substitute $(\bar{Z}Z-\bar{Z}\dot{Z})=-2ir^{2}\dot{\theta}$,
obtaining
\begin{dmath}
\dot{\theta}=\frac{C_{3}}{r^{2}}+\frac{km\lambda}{2}=\frac{k^{2}\lambda^{2}C_{3}}{4[\wp(\Omega)-\wp(t+\omega_{2})]}+\frac{km\lambda}{2},
\end{dmath}
where 
\begin{equation}
C_{3}\equiv k\left[\frac{m^{3}\lambda}{2}-C_{1}-m\lambda C_{2}\right]
\end{equation}
Using the identity
\begin{dmath}
\int\frac{dz}{\wp(z)-\wp(\Omega)}=\frac{1}{\wp'(\Omega)}\left[2z\zeta(\Omega)+\log\frac{\sigma(z-\Omega)}{\sigma(z+\Omega)}\right]
\end{dmath}
and taking $\theta(0)=0$, we have
\begin{eqnarray}
\theta(t)&=&\frac{k^{2}\lambda^{2}C_{3}}{4\wp'(\Omega)}\left[2t\zeta(\Omega)+\log\frac{\sigma(t+\omega_{2}-\Omega)\sigma(\omega_{2}+\Omega)}{\sigma(t+\omega_{2}+\Omega)\sigma(\omega_{2}-\Omega)}\right]\nonumber \\& &\;+\frac{km\lambda}{2}t
\end{eqnarray}

We can use the Weierstrass differential equation (\ref{eq:weierstrass_eqn})
directly to obtain $\wp'(\Omega)=(i/2)k^{2}\lambda^{2}C_{3}$, leading
to a seemingly remarkable cancellation. We then have 
\begin{multline}
e^{i\theta(t)}=\sqrt{\frac{\sigma(t+\omega_{2}+\Omega)\sigma(\omega_{2}-\Omega)}{\sigma(t+\omega_{2}-\Omega)\sigma(\omega_{2}+\Omega)}} \\ {\cdot \exp\left[-\left(\zeta(\Omega)+\frac{ikm\lambda}{2}\right)t\right]}.
\end{multline}
Finally, a few terms cancel in the overall expression for $Z$, yielding
\begin{multline}
Z(t)=\left[\frac{2}{\lambda k\sigma(\Omega)}\sqrt{\frac{\sigma(\omega_{2}-\Omega)}{\sigma(\omega_{2}+\Omega)}}\right]\frac{\sigma(t+\omega_{2}+\Omega)}{\sigma(t+\omega_{2})}\\ \cdot\exp\left[-\left(\zeta(\Omega)+\frac{ikm\lambda}{2}\right)t\right].
\end{multline}
\begin{figure}
\begin{minipage}[c][1\totalheight][t]{0.49\textwidth}%
\includegraphics[width=1\textwidth]{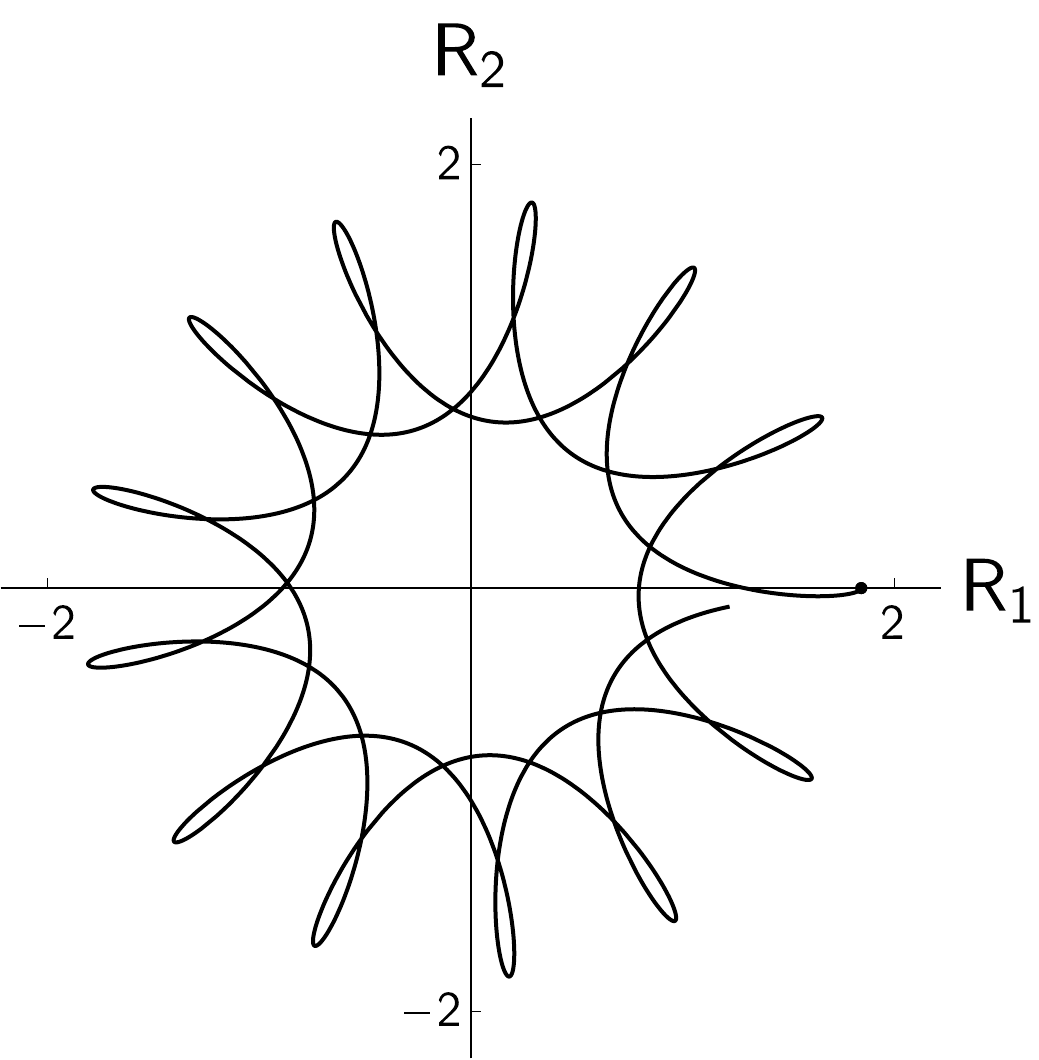}%
\end{minipage}\hfill{}%
\begin{minipage}[c][1\totalheight][t]{0.49\textwidth}%
\includegraphics[width=1\textwidth]{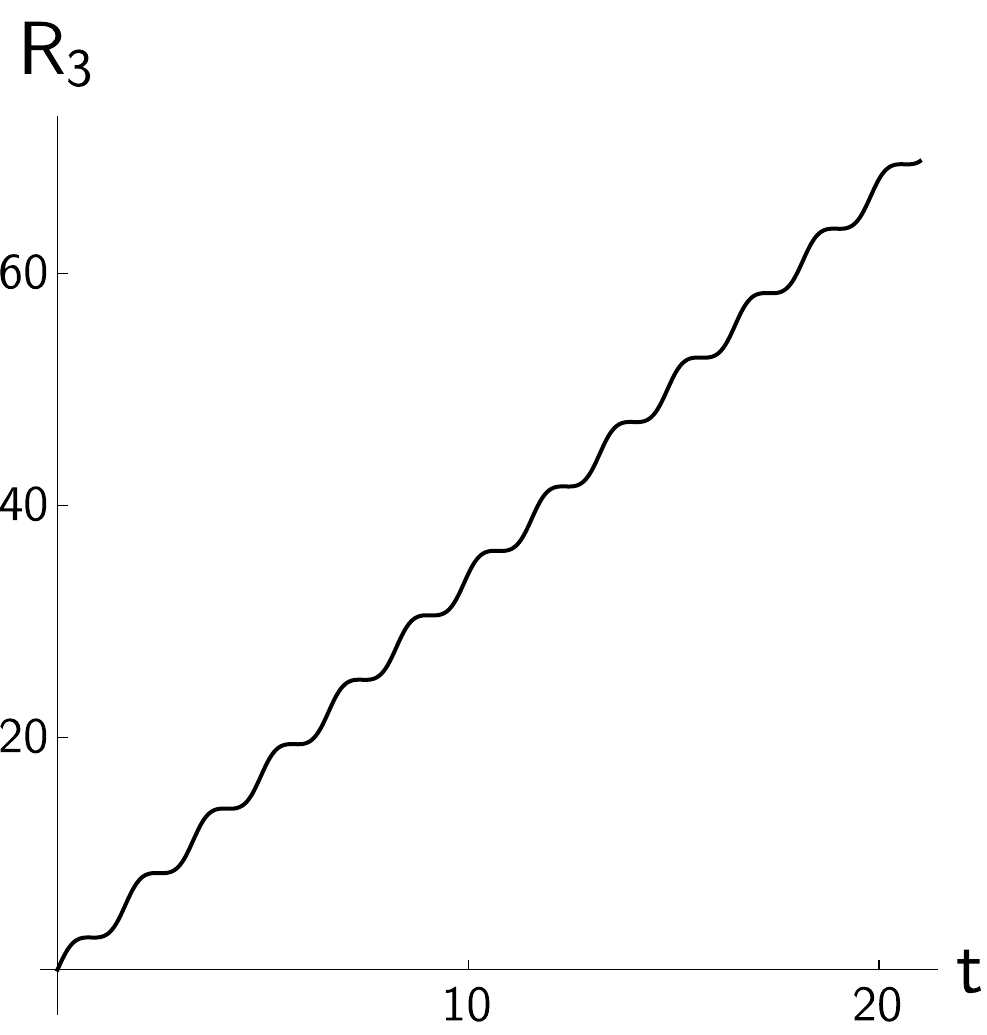}%
\end{minipage}

\caption{The orbit in the $R_{1}$-$R_{2}$ plane (above) and the evolution
of $R_{3}$ with time (below). The sample solution is plotted for
$0<t<21$ and uses parameters $k=1$, $\lambda=2.4$, $C_{1}=0.5$,
$C_{2}=1$, $s=2$, $m=0.5$. \label{fig:orbit_plot}}
\end{figure}

A sample solution is plotted in Fig. \ref{fig:orbit_plot}. We can
see that, in the $R_{1}$-$R_{2}$ plane, the solution traces an oscillating
curve in between some inner and outer radius. Meanwhile, the solution
propagates in the $R_{3}$ direction with non-uniform speed. This
behavior is typical over all parameter values we tested.

\section{Mechanical Interpretation and Quantization of the Reduced Systems \label{sec:reduced_system_quant}}

The equations of motion following from the ansatz (\ref{eq:eqmo_ansatz}),
defining the reduced system for $S_{1}$, can be written as
\begin{equation}
\ddot{R}=\lambda[\dot{R},[K,R]+mK]+[K,[K,R]].
\end{equation}
These are the equations of motion of a particle in a static electromagnetic
field, given by (working in cylindrical polar coordinates where $R_{1}=r\cos\theta,\;R_{2}=r\sin\theta,\;z=R_{3}$)
\begin{equation}
\vec{B}=kr\hat{\theta}+mk\hat{z},\quad\vec{E}=k^{2}r\hat{r},
\end{equation}
which follow from the vector and scalar potentials 
\begin{equation}
\vec{A}=\frac{\lambda k}{2}\left(mr\hat{\theta}+r^{2}\hat{z}\right),\quad V=\frac{k^{2}}{2}r^{2}.
\end{equation}
The classical Hamiltonian is then
\begin{equation}
H_{1}=\frac{1}{2}p_{r}^{2}+\frac{1}{2}\frac{\left[p_{\theta}-A_{\theta}\right]^{2}}{r^{2}}+\frac{1}{2}\left[p_{z}-A_{z}\right]^{2}+V(r).
\end{equation}

It is clear that $p_{\theta}$ and $p_{z}$ are conserved. This formulation lends some physical intuition to the solutions found in section \ref{sec:Reduction-mech}. We can
pass to the quantum theory as usual by finding the covariant Laplacian
in cylindrical co-ordinates,
\begin{multline}
\hat{H}_{1}\psi=-\frac{\hbar^{2}}{2}\frac{1}{r}\frac{\partial}{\partial r}\left[r\frac{\partial\psi}{\partial r}\right]+{\frac{1}{2r^{2}}\left[-i\hbar\partial_{\theta}-A_{\theta}\right]^{2}\psi}\\   +{\frac{1}{2}\left[-i\hbar\partial_{z}-A_{z}\right]^{2}\psi+V(r)\psi}.
\end{multline}
The conservation of $p_{\theta},\;p_{z}$ leads us to seek a solution
to the Schrodinger equation of the separable form
\begin{equation}
\psi(r,\theta,z)=\frac{1}{\sqrt{r}}\rho(r)e^{il\theta}e^{i\frac{p_{z}z}{\hbar}}
\end{equation}
for integer $l=\frac{p_{\theta}}{\hbar}$. The system is then reduced
to a one-dimensional Schrodinger equation
\begin{equation}
-\frac{\hbar^{2}\rho''(r)}{2}+U(r)\rho(r)=E\rho(r)
\end{equation}
with effective potential
\begin{equation*}
U(r)=-\frac{\hbar^{2}}{8r^{2}}+\frac{1}{2}\frac{\left[\hbar l-A_{\theta}\right]^{2}}{r^{2}}+\frac{1}{2}\left[p_{z}-A_{z}\right]^{2}+V(r)
\end{equation*}
\begin{multline}
=\frac{1}{2}\Biggl[\frac{\hbar^{2}\left[l^{2}-\frac{1}{4}\right]}{r^{2}}-\frac{\hbar k\lambda ml}{r}+\left(\frac{k^{2}\lambda^{2}m^{2}}{4}+p_{z}^{2}\right)\\ +(k-\lambda p_{z})kr^{2}+\frac{\lambda^{2}k^{2}r^{4}}{4}\Biggr].
\end{multline}

At low coupling $(\lambda\rightarrow0)$, we have 
\begin{equation}
-\hbar^{2}\rho''+\left\{ \frac{\hbar^{2}\left[l^{2}-\frac{1}{4}\right]}{r^{2}}+p_{z}^{2}+\frac{k^{2}}{2}r^{2}\right\} \rho=E\rho,
\end{equation}
and we see by dimensional analysis
\begin{equation}
[\hbar k]=1/L^{2}\quad\Rightarrow E\sim|\hbar k|.
\end{equation}
These are weakly coupled massless excitations. But in the high coupling
limit $(\lambda\rightarrow\infty)$, we have
\begin{equation}
-\hbar^2\rho''+\left\{ \frac{k^{2}\lambda^{2}}{4}(m^{2}+r^{4})\right\} \rho=E\rho,
\end{equation}
which yields a much more peculiar spectrum 
\begin{equation}
[\hbar^2 k\lambda]^{2}=1/L^{6}\quad\Rightarrow E\sim|\hbar^{2}\lambda k|^{2/3}.
\end{equation}
If this dispersion relation describes some fundamental constituents
of the theory, then they are certainly not particles in the traditional
sense. We propose that this may be a glimpse of some post-relativistic
constituents as mentioned in section \ref{sub:Relation-to-Slow}.

\subsection{quantization of the dual reduced system}

In the dual picture (nonlinear sigma model), our ansatz picks out
a class of solutions that correspond to a different mechanical system.
Though the equations of motion in each picture can be mapped to one
another via the duality, the correspondence is not immediately obvious,
and the systems will appear very different upon quantization.

After the ansatz, the duality relations (\ref{eq:dual}) read 
\begin{equation}
g^{-1}g'=\lambda e^{Kx}\left(S+\frac{1}{\lambda}K\right)e^{-Kx},\quad g^{-1}\dot{g}=\lambda e^{Kx}Le^{-Kx}.
\end{equation}

Writing $g=h(t,x)e^{-Kx}$ yields 
\begin{equation}
h^{-1}h'=\lambda S\quad h^{-1}\dot{h}=\lambda L.
\end{equation}
We further suppose that $h$ is separable as $h(t,x)=F(x)Q(t)$. Then
the equation for $S$ can be separated as 
\begin{equation}
F^{-1}(x)F'(x)=\lambda Q(t)S(t)Q^{-1}(t)\label{eq:sep}
\end{equation}
Both sides are equal to some constant traceless matrix $C$. Since
$Q(t)$ is only unique up to multiplication on the left by a constant
matrix in $\mathbb{SU}(2)$, we can use this to choose $C$ to be
diagonal and thus proportional to $K$. We then have
\begin{equation}
Q(t)S(t)Q^{-1}(t)=sK,
\end{equation}
implying that $\mathrm{Tr}S^{2}=s^{2}k^{2}.$ (\ref{eq:sep}) is satisfied
if 
\begin{equation}
F(x)=e^{\lambda sKx}.
\end{equation}
Thus, the full corresponding ansatz for the field variable in the
dual theory is
\begin{equation}
g(t,x)=e^{\lambda sKx}Q(t)e^{-Kx},
\end{equation}
where $Q$ is related to the previous variables by
\begin{equation}
S=sQ^{-1}(t)KQ(t),\quad L=\frac{1}{\lambda}Q^{-1}\dot{Q}.
\end{equation}
The dual Lagrangian can now be written as
\begin{equation}
\mathcal{L}_{2}=\frac{1}{2f^{2}}\mathrm{Tr}\left[\left(Q^{-1}\dot{Q}\right)^{2}-\left(\lambda sQ^{-1}KQ-K\right)^{2}\right].
\end{equation}
It is useful to parameterize $Q$ in terms of the Euler angles: 
\begin{equation}
Q=e^{\frac{i}{2}\sigma_{3}\gamma}e^{\frac{i}{2}\sigma_{1}\beta}e^{\frac{i}{2}\sigma_{3}\alpha}.
\end{equation}
The traces in $\mathcal{L}_{2}$ can then be computed directly, yielding
\begin{equation}
\mathcal{L}_{2}=\frac{1}{\lambda^{2}}\left\{ \frac{\dot{\alpha}^{2}+\dot{\beta}^{2}+\dot{\gamma}^{2}}{2}+\cos\beta\dot{\alpha}\dot{\gamma}-V(\beta)\right\} 
\end{equation}
where (dropping a constant shift)
\begin{equation}
V(\beta)=-2k^{2}\lambda s\cos\beta.
\end{equation}
As a mechanical system, this is the well known spinning top (isotropic
Lagrange top). It is instructive to write
\begin{equation}
\mathcal{L}_{2}=\frac{1}{\lambda^{2}}\left[\frac{1}{2}g_{ij}\dot{\alpha}^{i}\dot{\alpha}^{j}-V\right]
\end{equation}
where
\begin{equation}
g_{ij}=\left(\begin{array}{ccc}
1 & 0 & \cos\beta\\
0 & 1 & 0\\
\cos\beta & 0 & 1
\end{array}\right)
\end{equation}
is the metric of the rotation group and $V$ is the gravitational
potential of the top. The overall constant $\frac{1}{\lambda^{2}}$
in the action leads to a rescaling of $\hbar\mapsto\hbar\lambda^{2}$
upon quantization.

To pass to the quantum theory, we find the Laplacian operator with
respect to the metric $g$ of Eulerian coordinates, 
\begin{equation}
\nabla^{2}\psi=\frac{1}{\sqrt{g}}\partial_{i}\left[\sqrt{g}g^{ij}\partial_{j}\psi\right]
\end{equation}
The Hamiltonian is then
\begin{multline}
\hat{H}_{2}=-\frac{\hbar^{2}\lambda^{4}}{2}\biggl[\frac{\partial_{\alpha}^{2}+\partial_{\gamma}^{2}-2\cos\beta\ \partial_{\alpha}\partial_{\gamma}}{\sin^{2}\beta}\\+\partial_{\beta}^{2}+\cot\beta\partial_{\beta}\psi\biggr]+V\psi
\end{multline}
We can again reduce the Schrodinger equation $\hat{H}_{2}\psi=E\psi$
to a one-dimensional Schrodinger equation with the ansatz 
\begin{equation}
\psi(\alpha,\gamma,\beta)=e^{im_{\alpha}\alpha}e^{im_{\gamma}\gamma}\frac{B(\beta)}{\sqrt{\sin\beta}},
\end{equation}
yielding
\begin{equation}
-\hbar^{2}\lambda^{4}\frac{B''(\beta)}{2}+U(\beta)B(\beta)=EB(\beta)
\end{equation}
where
\begin{multline}
U(\beta)=-\frac{\hbar^{2}\lambda^{4}}{8}+\frac{\hbar^{2}\lambda^{4}}{2\sin^{2}\beta}\biggl[m_{\alpha}{}^{2}+m_{\gamma}^{2}\\-2m_{\alpha}m_{\gamma}\cos\beta-\frac{1}{4}\biggr]-2k^{2}\lambda s\cos\beta.
\end{multline}

This can be studied by standard techniques for periodic potentials
(Floquet theory or Bloch waves etc.) We content ourselves with a
quick look at low energy excitations: small oscillations around the
classical equilibrium $q=0$ and setting $m_{\alpha}=0=m_{\gamma}$.
Changing variables $\beta=\hbar\lambda^{2}q$ and expanding around
the classical minimum $q=0$ gives
\begin{dmath}
\shoveleft{-\frac{1}{2}\frac{d^{2}B}{dq^{2}}+\left\{ q^{2}\left(\hbar^{2}k^{2}s\lambda^{5}\right)-\frac{1}{8q^{2}}-2k^{2}s\right\} B\approx EB\;\;}
\end{dmath}
The solutions involve Laguerre polynomials and the spectrum is, in
this approximation $E_{n}\approx\sqrt{2}(2n+1)\hbar k\sqrt{s}\lambda^{\frac{5}{2}}.$
If we remove the zero-point energy ($n=0$), we have the energy of
$n$ free particles each of energy $e_{1}=\hbar k\sqrt{8s}\lambda^{\frac{5}{2}}$. This is the dispersion relation of massless particles, except for
a rescaling of the speed. $ \\ $

\section{Conclusions and outlook\label{sec:Conclusion-and-future}}

Because they only exist in the short distance limit, it is difficult to say whether objects like ``preons'' we discuss could correspond to directly
observable objects in an experiment. Quarks were not considered at
first to be directly observable things either, as they could not be
created as isolated particles. In the $S_1$-model's strong coupling limit, the Minkowski
geometry of space-time appears to be lost, and wave propagation is
sustained entirely by the non-linearity. However, these waves do not appear to transmit information, and perhaps any ``post-relativistic''
effects are hidden by some sort of confinement when
they form bound states. 

It is at least intriguing to question whether highly coupled theories
have fundamental constituents whose nature is so exotic that they
have been overlooked. Drawing paralells with $\lambda\phi^{4}$ theory,
it is tempting to speculate that the Higgs particle of the standard
model is such a composite of some strongly bound preons existing only
at short distances. Were this the case, one could sensibly describe
a ``pure Higgs'' at short distances.

For a more complete understanding, we must quantize the whole theory
rather than just its mechanical reduction. Since the equations have
a Lax pair, it should be possible to perform a canonical transformation
to angle-variables and then pass to the quantum theory. Such a quantization
was achieved for sine-Gordon theory \cite{QuantThSolitons}, proving
that the solitons are fermions which bind to form the scalar waves.
A similar analysis of our model is a lengthy endeavor, and we hope
to return to this later after laying the groundwork and motivation
here. 

We present a side by side comparison of comparison of our work with the two models in Table \ref{tab:thetable} below. 

\section{Acknowledgements}

S.G.R. would like to thank Govind Krishnaswami and V.V. Sreedhar for discussions. The work of E.R. is supported by the NSF Graduate
Research Fellowship Program under Grant No. DGE-1419118 and by the University of Rochester Sproull Fellowship.

\renewcommand{\arraystretch}{1.75}

\begin{center}
\vspace{0pt}
\end{center}
\begin{widetext}
\begin{center}
\begin{table}
\begin{tabular}{|c|c|c|}
\multicolumn{1}{c}{Nilpotent Field Theory ($S_{1}$) } & \multicolumn{1}{c}{} & \multicolumn{1}{c}{Principal Chiral Model ($S_{2}$)}\tabularnewline
\hline 
$\mathcal{L}_{1}=\mathrm{Tr}\left\{ \frac{1}{2\lambda}\dot{\phi}^{2}-\frac{1}{2\lambda}{\phi'}^{2}+\frac{1}{3}\phi[\dot{\phi},\phi']\right\} $  & Lagrangian density  & $\mathcal{L}_{2}=\frac{1}{2\lambda^{2}}\mathrm{Tr}\left\{ (g^{-1}\dot{g})^{2}-(g^{-1}g')^{2}\right\} $\tabularnewline
\hline 
$I=\frac{1}{\lambda}\dot{\phi},\quad J=\phi'$  & Currents  & $I=g^{-1}g',\quad J=g^{-1}\dot{g}$\tabularnewline
\hline 
$\lambda I'=\dot{J}$  & Current Identity  & $\dot{I}-\frac{1}{\lambda}J'+\lambda\left[J,I\right]=0$\tabularnewline
\hline 
$\dot{I}-\frac{1}{\lambda}J'+\lambda\left[J,I\right]=0$  & Equation of Motion  & $\lambda I'=\dot{J}$\tabularnewline
\hline 
\multicolumn{1}{c}{Reduced System of $S_{1}$} & \multicolumn{1}{c}{} & \multicolumn{1}{c}{Reduced System of $S_{2}$}\tabularnewline
\hline 
$\phi(t,x)=e^{Kx}R(t)e^{-Kx}+mKx$  & Wave Ansatz  & $g(t,x)=e^{\lambda sKx}Q(t)e^{-Kx}$\tabularnewline
\hline 
$I=e^{Kx}\dot{R}e^{-Kx}$  & Wave Currents  & $I=e^{Kx}\left\{ \lambda sQ^{-1}KQ-K\right\} e^{-Kx}$\tabularnewline
$J=e^{Kx}\left\{ \lambda[K,R]+mK\right\} e^{-Kx}$  &  & $J=e^{Kx}\bigl\{ Q^{-1}\dot{Q}\bigr\} e^{-Kx}$\tabularnewline
\hline 
$S=\dot{R}+\frac{1}{\lambda}K$  & Common Variables  & $S=sQ^{-1}KQ$\tabularnewline
$L=[K,R]+mK$  &  & $L=\frac{1}{\lambda}Q^{-1}\dot{Q}$\tabularnewline
\hline 
$\dot{L}=[K,S]$  & Current Identity  & $\dot{S}=\lambda[S,L]$\tabularnewline
$\dot{S}=\lambda[S,L]$  & Eqn. of motion  & $\dot{L}=[K,S]$\tabularnewline
\hline 
$H_{1}=\mathrm{Tr}\left\{ \frac{1}{2}\left(S-\frac{1}{\lambda}K\right)^{2}+\frac{1}{2}L^{2}\right\} $  & Hamiltonian $H_{1}=H_{2}$  & $H_{2}=\mathrm{Tr}\left\{ \frac{1}{2}\left(S-\frac{1}{\lambda}K\right)^{2}+\frac{1}{2}L^{2}\right\} $\tabularnewline
\hline 
$\mathcal{L}_{1}=\mathrm{Tr}\Bigr\{\frac{1}{2}\left(S-\frac{1}{\lambda}K\right)^{2}\quad\quad\quad\quad\quad$  & Lagrangian $\mathcal{L}_{1}\neq\mathcal{L}_{2}$  & $\mathcal{L}_{2}=\frac{1}{2}\mathrm{Tr}\left\{ L^{2}-\left(S-K\right)^{2}\right\} $\tabularnewline
$\quad\quad\quad\quad\quad+\frac{1}{3}R\left[S-K,L\right]-\frac{1}{2}L^{2}\Bigr\}$ &  & \tabularnewline
\hline 
$E\sim|k|^{2/3}\quad(\lambda\rightarrow\infty)$  & short-range dispersion  & $E\sim|k|\quad(\lambda\rightarrow0)$\tabularnewline
\hline 
\end{tabular}
\caption{\label{tab:thetable} A comparison of results in the dual models}
\end{table}
\end{center}
\end{widetext}

\appendix

\section{quadratic hamiltonians with nilpotent bracket algebras \label{sec:Appendix:nilpotent}}

Many classical systems have a quadratic Hamiltonian together with Poisson brackets (or commutators in the
quantum case) which generate a Lie algebra:
\begin{equation}
H=\frac{1}{2}h^{ab}v_{a}v_{b}, \quad\quad \left\{ v_{a},v_{b}\right\} =c_{ab}^{c}v_{c}
\end{equation}

Where $c_{ab}^{c}$ are the structure constants of the bracket algebra
and $v_{a}$ the dynamical variables. The quadratic nature of the
Hamiltonian immediately affords a geometric interpretation: $h_{ab}$
defines a left-invariant metric, $h\in \mathfrak{g}\vee\mathfrak{g}$
on the Lie group $G$ (with generating algebra $\mathfrak{g}$). The equations of motion then describe
geodesics on the Lie group under this metric.

A nilpotent Lie Algebra of step $n$ is a Lie algebra in which all
repeated brackets of order $n$ vanish. The combination of a quadratic
Hamiltonian \emph{and }a nilpotent bracket algebra can allow one to
solve for the spectrum of a quantum system algebraically, using only
the representation structure of the associated Lie group. This is
done by using a representation to generate raising and lowering operators,
as is familiar in the case of the harmonic oscillator. While we did
not take this (somewhat ambitious) approach here, it is carried out
in \cite{JorgensenKlink1} for a magnetic system very similar to the
one we discuss in section \ref{sec:reduced_system_quant}. It is worth
at least mentioning this point of view, as it connects the model studied
here to other, more well-known models.

\subsection{Nilpotent Mechanical Systems}

The simplest system of this type is a harmonic oscillator, 
\begin{equation}
H=\frac{1}{2}\left(p^{2}+\omega^{2}q^{2}\right),\quad\quad
\left\{ p,q\right\} =1.
\end{equation}

Here the canonical variables $p$ and $q$ form a step-2 nilpotent
Lie algebra, where all double commutators vanish. Of course, any Hamiltonian
in terms of the canonical variables will have this bracket algebra,
but what happens in the non-quadratic case? Consider the anharmonic
oscillator, 
\begin{equation}
H=\frac{1}{2}\left(p^{2}+\omega^{2}q^{2}\right)+\lambda q^{4}
\end{equation}

We can recast this as a quadratic Hamiltonian with step-3 nilpotent
bracket algebra by defining $q_{2}=q^{2}$, and then treating this
as a distinct element of the algebra. We then have 
\begin{equation}
H=\frac{1}{2}\left(p^{2}+\omega^{2}q^{2}\right)+\lambda q_{2}^{2}
\end{equation}
\begin{equation}
\left\{ p,q_{2}\right\} =2q,\quad\left\{ p,q\right\} =1,\quad\left\{ q_{2},q\right\} =0,
\end{equation}
where one can see that all triple commutators vanish. Thus, the classical
anharmonic oscillator describes geodesics in the corresponding nilpotent
Lie group. It is then possible to solve such quantum theories using the methods of \cite{JorgensenKlink1}.

The mechanical reduction of our field theory gives another example.
The equations of motion (\ref{eq:eqmo_ansatz}) follow from the hamiltonian
(\ref{eq:Ham_Ansatz}) with Poisson Brackets
\begin{dmath}[compact]
\left\{ S_{a},S_{b}\right\} =\lambda\epsilon_{abc}L_{c},\quad\left\{ L_{a},L_{b}\right\} =0,\quad\left\{ S_{a},L_{b}\right\} =\epsilon_{abc}K_{c}
\end{dmath}

This a step three nilpotent Lie algebra. Its representation on the space of functions
of $\mathbb{R}$ is what we used to quantize the theory. In the dual picture
the Poisson brackets would not be nilpotent: the Lie algebra it is
the semi-direct product of $SU(2)$ with an abelian algebra.

\subsection{Nilpotent Field Theories}

Interestingly, we find that some of the most important field theories
of particle physics can be described in this way. The bosonic part
of the standard model consists of Yang-Mills theory coupled to a Higgs
sector described as a $\lambda\phi^{4}$ scalar field theory.

In field theory, the obvious analogue of the anharmonic oscillator
is pure $\lambda\phi^{4}$ theory. Identifying $\phi_{2}(x)=\phi^{2}(x)$,
this theory can be described by
\begin{equation}
H=\frac{1}{2}\int\left[\pi^{2}(x)+m^{2}\phi^{2}(x)+\lambda\phi_{2}^{2}\right]dx
\end{equation}
\begin{dmath}[compact]
\left\{ \pi(x),\phi_{2}(y)\right\} =2\phi(x)\delta(x-y),\quad\left\{ \pi(x),\phi(y)\right\} =\delta(x-y)
\end{dmath}

\noindent Pure $\lambda\phi^{4}$ theory (without coupling to fermions) in 4
dimensions remains intractable in the short distance limit: is not
an asymptotically free theory. We see here that it follows from a
Hamiltonian with one degree of extra nilpotency in the bracket algebra.
This suggests that perhaps the theory is easier tamed with an algebraic
approach.

Another famously puzzling theory, Yang-Mills theory, can be cast in
the same language. Here the Poisson Brackets and Hamiltonian are best
expressed in terms of the electric field
\begin{equation}
E[a]=\int E^{bi}a_{bi}dx
\end{equation}
(where $a$ is a smooth test function) and the magnetic field $B=dA+A\wedge A$:
\begin{eqnarray}
\nonumber &\{E[a],B\}&=da+[A,a],\\ \nonumber &\{E[a],A\}&=a,\\ &\{A_{aj}(x)_{,}A_{bj}(y)\}&=0,
\end{eqnarray}
\begin{equation}
 H=\frac{1}{2}\int\left(E^{2}+B^{2}\right)dx.
\end{equation}
Yang Mills theory, however, is an asymptotically free theory. The
fact that it can be brought to the same form as pure $\lambda\phi^{4}$
theory suggests some commonality in structure of the two theories,
though they might appear glaringly different due to their short-distance
behavior.

We pause to note that not all systems are nilpotent. The simplest
example would be the rigid rotor, whose bracket algebra is that of
angular momentum, where repeated commutators do not vanish. Such a
Lie algebra is perhaps misleadingly labeled as \emph{simple} in the
mathematics literature. Also, the Euler equations of an ideal fluid
can be formulated with a quadratic hamiltonian on the Lie algebra
of vector fields. Nilpotent Lie algebras could be useful as approximations
here.

\subsection{The current algebra of $S_{1}$}

The equations of motion (\ref{eq:eqmo}) follow from the hamiltonian
\begin{equation}
H_{1}=\frac{1}{2}\int[\lambda I_{a}I_{a}+\frac{1}{\lambda}J^{a}J^{a}]dx
\end{equation}
and the Poisson brackets from $S_{1}$,
\begin{eqnarray}
\nonumber
&\{J^{a}(x),\ J^{b}(y)\}_{1}&=0\\ \nonumber &\{I_{a}(x),\ J^{b}(y)\}_{1}&=-\delta_{a}^{b}\delta'(x-y) \\ &\{I_{a}(x),\ I_{b}(y)\}_{1}&=\epsilon_{abc}J^{c}\delta(x-y).
\end{eqnarray}

So this theory can also be cast as a quadratic Hamiltonian with step-3
nilpotent algebra. This further motivates the analogy between our
model and $\lambda\phi^{4}$ theories.

It is natural, in nilpotent Lie algebras, to take the singular limit
of the metric where the coefficient of the higher-step generators
shrinks to zero. (This geometry has been well-studied in the simplest
case of the Heisenberg group \cite{subRiemannian}). This is precisely
the strong coupling limit $\lambda\to\infty$ of our theory: the second
term in the hamiltonian $\frac{1}{2}\int[\lambda I_{a}I_{a}+\frac{1}{\lambda}J^{a}J^{a}]dx$
tends to zero. In this limit the co-metric is not invertible.

The resulting sub-Riemannian geometry still has geodesics connecting nearby points: the Hormander condition is satisfied because the commutator of the surviving generators $I_a(x)$ generate the remaining ones $J_a$. The Chow-Rashevsky theorem does not directly apply here as we are dealing with an infinite dimensional manifold. But, it does suggest that there are propagating solutions even in the limit $\lambda\to \infty$. We found some examples numerically first, and then  found analytic solutions including these examples. So at least in this case, the intuition provided by the sub-Riemannian geometry was useful in understanding the strong coupling limit.

\end{document}